\documentclass[prb,twocolumn,amssymb,amsmath,notitlepage,floatfix]{revtex4-1}
\usepackage{graphicx,subfigure}
\usepackage{amsfonts}
\usepackage{verbatim}
\usepackage{color}
\usepackage{epstopdf}
\usepackage{hyperref}
\begin{document}
	\title{ Enhanced current rectification in graphene nanoribbons:\\ Effects of geometries and orientations of nanopores } 
	\author{Joydeep Majhi}
	\email{joydeepmjh@gmail.com}
	\affiliation{Physics and Applied Mathematics Unit, Indian Statistical
		Institute, 203 Barrackpore Trunk Road, Kolkata-700 108, India}

	\author{Sudin Ganguly}
	\email{sudinganguly@gmail.com}
	\affiliation{Department of Physics, School of Applied Sciences, University of Science and Technology, Techno City, Kiling Road, Baridua 9th Mile, Ri-Bhoi, Meghalaya-793 101, India}

	\author{Santanu K. Maiti}
	\email{santanu.maiti@isical.ac.in}
	\affiliation{Physics and Applied Mathematics Unit, Indian Statistical
  Institute, 203 Barrackpore Trunk Road, Kolkata-700 108, India}

\begin{abstract}
    We discuss the possibility of getting rectification operation in graphene nanoribbon (GNR). For a system to be a rectifier, it must be physically asymmetric and we induce the asymmetry in GNR by introducing nanopores. The rectification properties are discussed for differently structured nanopores. We find that shape and orientation of the nanopores are critical and sensitive to the degree of current rectification. As the choice of Fermi energy is crucial for obtaining significant current rectification, explicit dependence of Fermi energy on the degree of current rectification is also studied for a particular shape of the nanopore. Finally, the role of nanopore size and different spatial distributions of the electrostatic potential profile across the GNR are discussed. Given the simplicity of the proposed method and promising results, the present proposition may lead to a new route of getting current rectification in different kinds of materials where nanopores can be formed selectively. 
\end{abstract}

\maketitle

\section{Introduction}
Diodes are the most basic electrical components that may be utilized to design logical circuits and memory components on a nanoscale \cite{konenkamp85}. Following Aviram and Ratner's \cite{aviram29} groundbreaking work on molecular rectifier, the construction of an efficient rectifier at the nanoscale is now the focus of intense research due to its economic need, affordability, and speed. Additionally, this leads to a new age of electron transport via a single molecule connected to two nanocontacts, which is the fundamental mechanism of molecular electronics. Recent advances in experimental techniques for manipulating and contacting individual molecules, as well as the availability of leading-edge methods for accurately describing the electrical properties of single molecular junctions at the atomic level, have accelerated the study of electron transport through nano junctions \cite{dhirani106, zhou71, frantti59, metzger32, taylor89, ashwell126, perez1, nijhuis132, yee5, batra13, yoon136, wang141}.

The key concept for a conductor to be a rectifier is that it must be physically asymmetric, such that the Hamiltonian describing the system under forward and reverse biases become different, leading to an asymmetric charge current, that is $I(V) \neq I(-V)$ via the channel. The unbalanced current for two biased conditions can be accomplished in two ways: (i) by making an asymmetric conductor with respect to the source and drain electrodes~\cite{zhou71, mujica104, mujica281, krzeminski64, kornilovitch66}, or (ii) by making asymmetric interfaces between the conductor and electrodes~\cite{ zahid70, metzger212, dalgleish73}. In the presence of a finite bias, due to the aforementioned asymmetries, the resonant energy levels organize differently for two biased conditions. Consequently, charge transfer happens in one favored direction, resulting in current rectification. 

So far, various simple and complex molecular structures have been investigated theoretically for modulating the rectification performance. For instance, molecular wires~\cite{zhou71}, complex molecular structure with metallic electrodes~\cite{krzeminski64}, quantum wires with correlated site potentials\cite{saha93,patra484}, DNA molecular system with Fibonacci sequence~\cite{saha52}, etc. The rectification behavior has also been tested experimentally for several molecular systems, such as symmetric tetraphenyl and non-symmetric diblock dipyrimidinyldiphenyl molecules~\cite{perez1}, molecular rods~\cite{melbing}, single melamine molecule adsorbed on a Cu(100) surface~\cite{span}, etc. Several rectifiers based on DNA have also been proposed and experimentally validated \cite{guo86,guo8}. Despite the significant advances of molecular rectifiers since its proposal by Aviram and Ratner, the rectification efficiencies of fabricated molecular diodes are still very low compared to those of commercial silicon-based diodes. The limited success is mostly due to the instability of the active molecular structures, complicated fabrication procedures, and expensive measurement techniques, etc~\cite{rev-mol}.

On the other hand, graphene has the potential to be a substantial substitute for the aforementioned systems. Graphene, a single layer of carbon atoms organized in a two-dimensional honeycomb structure, has exceptional electrical, mechanical, thermal, and optoelectronic properties, as well as very high carrier mobility and long carrier mean free path longer than $1\mu m$ at room temperature \cite{neto81, liao15, schwierz101}. Such intriguing characteristic features have sparked a growing interest in graphene as a potential electronic material. Unfortunately, the absence of a bandgap in graphene limits its possibility in practical applications. Interestingly, finite stripes of graphene, namely, graphene nanoribbons (GNRs) show band gap openings~\cite{han-prl}, thereby, unleashing a wide range of possibilities in nano-electronics including rectification.

\begin{figure*}[ht]
		\includegraphics[width=1\textwidth]{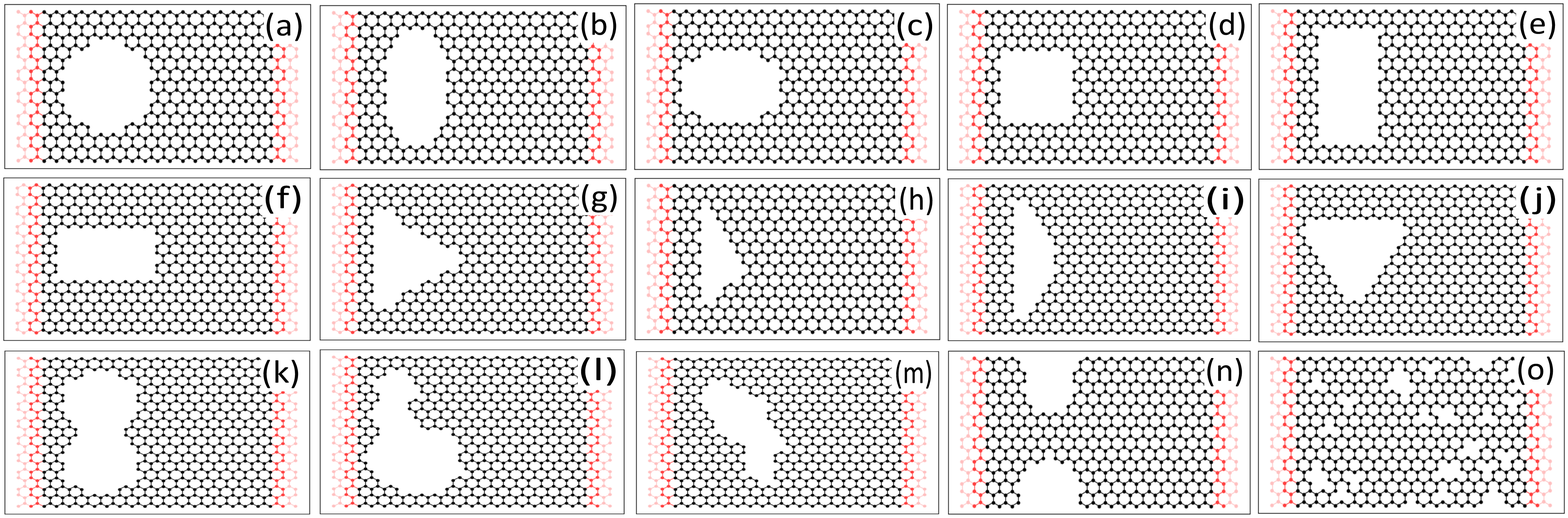}
	\caption{(Color online). Schematic illustration of graphene nanoribbons with various nanopore geometries. The geometries of the nanopores are (a) circle, (b) ellipse-I: semi-major axis along the width of the ZGNR, (c) ellipse-II: semi-major axis along the length of the ZGNR, (d) square, (e) rectangle-I: larger side is along the width of ZGNR, (f) rectangle-II: larger side is along the length of the ZGNR, (g) isosceles triangle (AAA), (h) scalene triangle (AAZ), (i) isosceles triangle (AZZ), (j) equilateral triangle (ZZZ), (k) arbitrary shape-I, (l) arbitrary shape-II, (m) arbitrary shape-III, (n) arbitrary shape-IV, and (o) random cavities. Here Z: side of the triangle with zigzag shape, A: edge of the triangle with armchair shape.}
	\label{fig1}
\end{figure*}

In recent years, various GNR-based structures have been explored with high-performance rectification properties. For instance, a structure consisting of GNR and single-walled carbon nanotube has been experimentally realized~\cite{dwei}, which exhibits a strong gate-dependent rectifying behavior~\cite{dwei,xhzang}. Z-shaped GNR junction having two armchair leads and a zigzag junction show rectification in presence of an external gate voltage~\cite{fwang}. A fused structure of an armchair GNR (AGNR) and a zigzag  GNR (ZGNR)~\cite{xfli}, edge hydrogenated ZGNR~\cite{jzeng}, a triangular-shaped zigzag nanoflake sandwiched between two asymmetric ZGNR electrodes~\cite{xlji}, GNRs doped with different atoms~\cite{tchen,lxie,xh-chem,eszam} also show good rectification performances. All the works mentioned here have exploited certain asymmetric configurations, that are either derived from the active region or the asymmetric source and drain electrodes. A similar kind of asymmetry can also be induced by drilling nanopores in GNRs and may have rectification properties. Such an attempt on rectification has not been done so far to the best of our concern.

Drilling nanopores in GNR requires sophisticated manufacturing procedures. Experimentally, drilling an array of nanopores in GNR has just been accomplished\cite{bai5,tada107}. Furthermore, Michael\cite{Fischbein93} and his co-workers created nanometer-scale pores in graphene by controlled exposure to a concentrated electron beam from a transmission electron microscope, and most significantly they do not evolve over time. Nanolithographic techniques\cite{zhang6}, template growth approaches\cite{safron8}, and Helium Ion beam milling\cite{zhang6,kalhor114,abbas8}, to mention a few, may all be used to get nanopores with different shapes and precise sizes. In light of the possibility to create nanopores with different shapes and sizes, it demands a thorough analysis of the rectification properties of GNRs with nanopores. Motivated by the preceding discussions, we conduct an in-depth investigation of the structural impacts of nanopores in ZGNR in order to achieve a high rectification ratio. Several regular and irregular geometries of nanopores are addressed here, including circles, ellipses, rectangles, and triangles, among others (Fig.~\ref{fig1}). The underlying physical mechanism of getting current rectification relies on the difference in transmission spectra in two different bias polarities. In most of the studies, the difference is caused in presence of impurities. But, for our present analysis, we do not need to consider any kind of such impurities either in site energies or in hopping integrals. We hope that our proposition leads to a new route of getting current rectification in different kinds of materials where cavities can be formed selectively. 

The key findings of the present work are: (i) asymmetrically placed nanopores exhibit finite rectification, (ii) a very high degree of rectification is achieved for a particular nanopore shape, and (iii) the degree of rectification can be tuned by varying the size of the nanopores.

The rest of the paper is organized as follows. In Section II, we present our model with differently structured nanopores and the theoretical formulae. In Sec. III, we discuss the structural impact of nanopores on the rectification properties. Finally, in Section IV, we summarise our key results.

\section{Model and Theoretical Formulation}

Figure~\ref{fig1} shows the schematic representations of different possible geometries of the nanopores (NPs) on the ZGNR that are examined in the present work. Two identical leads, denoted with red color, are symmetrically attached on either side of the ribbon. The carbon atoms in the central scattering region are represented by the black color. We consider several regularly shaped NPs, such as circular, elliptical, square, rectangular, different triangular NPs (Fig.~\ref{fig1}(a-j)), as well as some irregularly shaped NPs (Figs.~\ref{fig1}(k-m)). In addition to these regularly shaped NPs, a ZGNR structure is considered, which looks like a quantum dot (Fig.~\ref{fig1}(n)) and also a ZGNR with several small irregularly shaped NPs (absence of one or two carbon atoms: random cavities) (Fig.~\ref{fig1}(o)). In all the schematics of Fig.~\ref{fig1}, the structural asymmetry can readily be noticed with respect to the center of the central scattering region.

We describe our system in the nearest-neighbor tight-binding framework, and the corresponding Hamiltonian reads as
\begin{equation}
    H = \sum_{i}\epsilon_i c_i^{\dagger}c_i + \sum_{\langle ij\rangle} t \left(c_i^{\dagger}c_j + h.c. \right).
\end{equation}
$c_i^{\dagger}$ $(c_i)$ symbolizes electronic creation (annihilation) operator. $\epsilon_i$ is the on-site potential at the $i$-th site and $t$ is the nearest-neighbor hopping (NNH) integral.

\begin{figure}[ht]
	\includegraphics[width=0.45\textwidth]{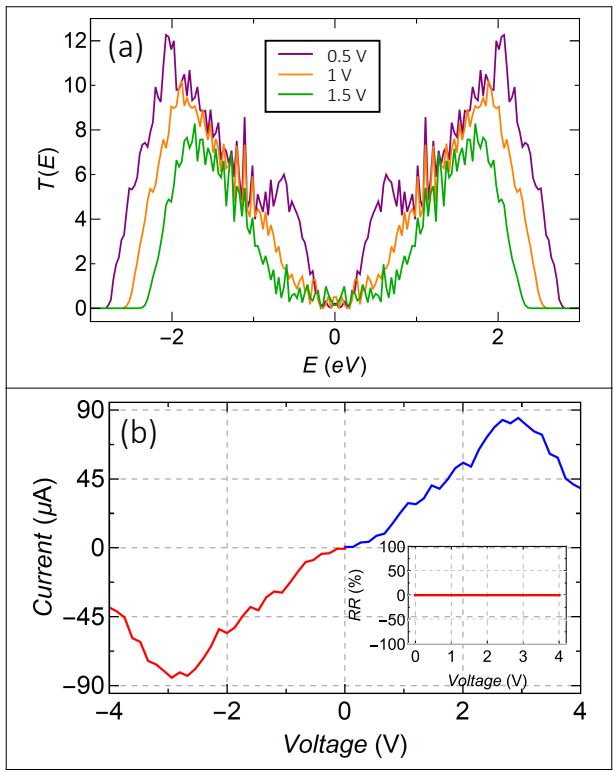}
	\caption{(Color online). Behavior of transmission probability as a function of energy at voltages 0.5$\,$V (violet color), 1$\,$V (orange color), and $1.5\,$V (green color) for forward biased condition. The results for the reversed biased cases are identical with that for the forward biased cases. (b) $I$-$V$ characteristics. Blue and red colors denote the currents in the forward and reverse biased conditions, respectively.}
	\label{fig2}
\end{figure}

\begin{figure*}[ht]
	\includegraphics[width=1\textwidth]{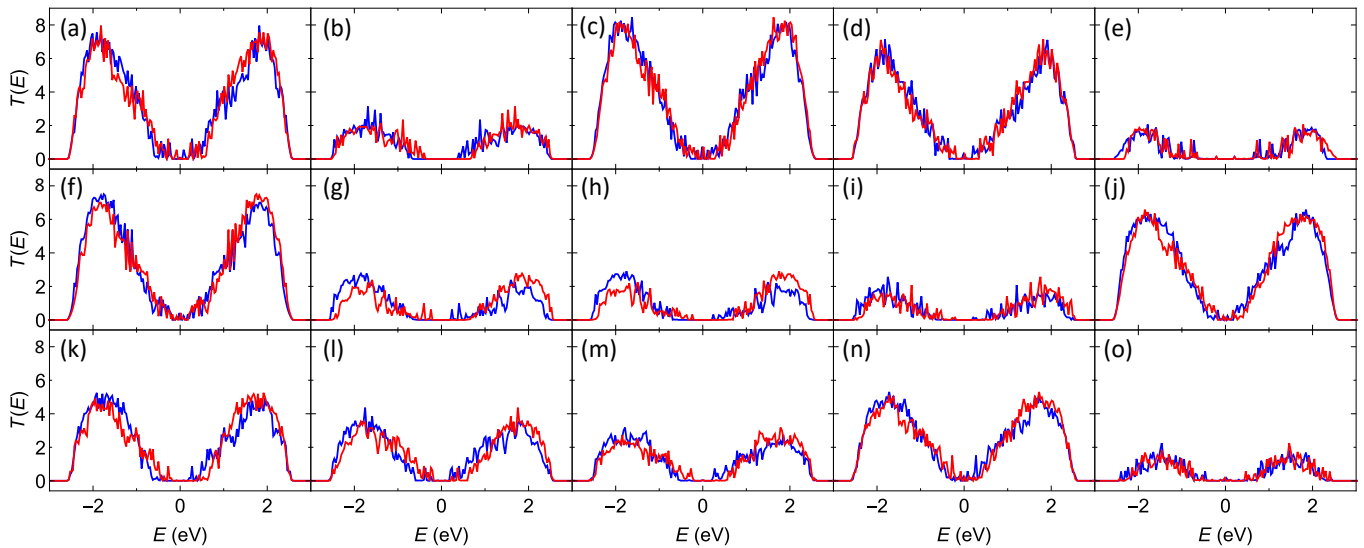}
	\caption{(Color online). Behavior of transmission probability as a function of energy corresponding to the nanopore shapes described in Fig.~\ref{fig1}. The ordering of the results are similar to that of Fig.~\ref{fig1}. Blue and red colors denote the transmission probabilities in the forward and reverse biased conditions, respectively.}
	\label{fig3}
\end{figure*}

In the presence of a finite bias voltage $V$, an electrostatic field develops across the scattering region, between source and drain, and therefore its site energies become voltage dependent~\cite{pleutin118,maiti377} while the site energies of the leads are unaffected. The actual modification of this potential profile is extremely difficult to predict since it falls under a complete many-body problem. However, such a scenario can be included `phenomenologically' into the system Hamiltonian, as previously described in other studies, by taking into account distinct types of potential profiles. Different kinds of potential profiles have been used in literature, out of which the linear dependence is the most common one~\cite{pleutin118,maiti377,cguo,msaha}. To include such kinds of potential profiles, the on-site potential term needs to be split up into two parts: one is voltage-independent and another is voltage-dependent. The total on-site potential at site $i$ can be expressed as 
\begin{equation}
\epsilon_i = \epsilon_i^0 + \epsilon_i(V). 
\end{equation}
$\epsilon_i^0$s depends on the nature of the material in use, could be disordered randomly or deterministically as well. The voltage-dependent potential $\epsilon_i(V)$ is connected to the electron screening as well as the bare electric field at the junction. In the absence of a screening electric field, $\epsilon_i(V)$ will have a linear distribution across the length of the active region, while the lateral voltage profile will be unchanged. As different materials have different electron screening and hence the different field variations, in the present work, we take into account three different potential profiles to get a complete picture of current rectification in presence of bare and screened electric field profiles as well. 

The two-terminal transmission probability is evaluated using the package KWANT, where transmission probabilities can easily be computed for different complicated lattice structures by connecting them with finite width leads. KWANT evaluates the transmission probability using the wave function approach based on the scattering matrix formalism~\cite{etms,qtat}. The scattering matrix formalism is mathematically equivalent to the non-equilibrium Green's function technique due to the Fisher-Lee relation~\cite{Fisher-Lee}, where the effects of the leads are taken through the self-energy terms and are assumed as a superposition of plane waves. After obtaining the propagating modes, the tight-binding equations $H\psi_i = \epsilon\psi_i$ are solved to get the scattering matrix, where $\psi_i$ is the wave function inside the scattering region and $H$ is the total Hamiltonian of the system in the tridiagonal block form which includes the Hamiltonian matrix of the scattering region, the Hamiltonian of the unit cell of the leads and other relevant block matrices. Finally, the transmission probability is calculated as $$T(E) = \sum\limits_{a\in p, b\in q} \lvert S_{pq}\rvert^2,$$ where $a$ and $b$ refer to the left and right electrodes. For a detailed discussion see Ref.~\cite{kwant}.

The net junction current for a given bias voltage $V$ at absolute zero temperature can be computed using the relation,
\begin{equation}
    I(V) = \frac{2e}{h}\int\displaylimits_{E_F - \frac{eV}{2}}^{E_F + \frac{eV}{2}} T(E) ~ \mathrm{d}E,
\label{curr}
\end{equation}
where $E_F$ represents the Fermi energy. 

Finally, the rectification ratio is defined as
\begin{equation} 
RR = \left| \frac{\mid I(+V)\mid - \mid I(-V)\mid}{\mid I(+V)\mid + \mid I(-V)\mid} \right| \times 100\%.
\end{equation}
 $RR = 0$ does not imply a rectification. On the other hand, $RR=100\%$ signifies maximum current rectification. Our ultimate goal is of course to have high degree of current rectification as much as it is possible.

\section{Numerical results and discussion}
Before we discuss the results, let us first mention the different parameters and units used in the present work. All the energies are measured in electron-volt (eV). For the sake of simplicity, the on-site potential $\epsilon_i^0$ and the on-site potential for the leads are set at zero. In the active region, the NNH integral is fixed at $t = 1\,$eV, while for the leads it is 2$\,$eV, to work within the wide-band limit. The length and width of the central scattering region are about 15$\,$nm, and 10$\,$nm, respectively. The widths of the left and right leads have the same lateral dimensions as that of the central scattering region. Unless otherwise stated, all the currents are computed by setting the Fermi energy at $E_F=0.5\,$eV and the field dependency is assumed to be linear across the active region.

\begin{figure*}[t]
	\includegraphics[width=1\textwidth]{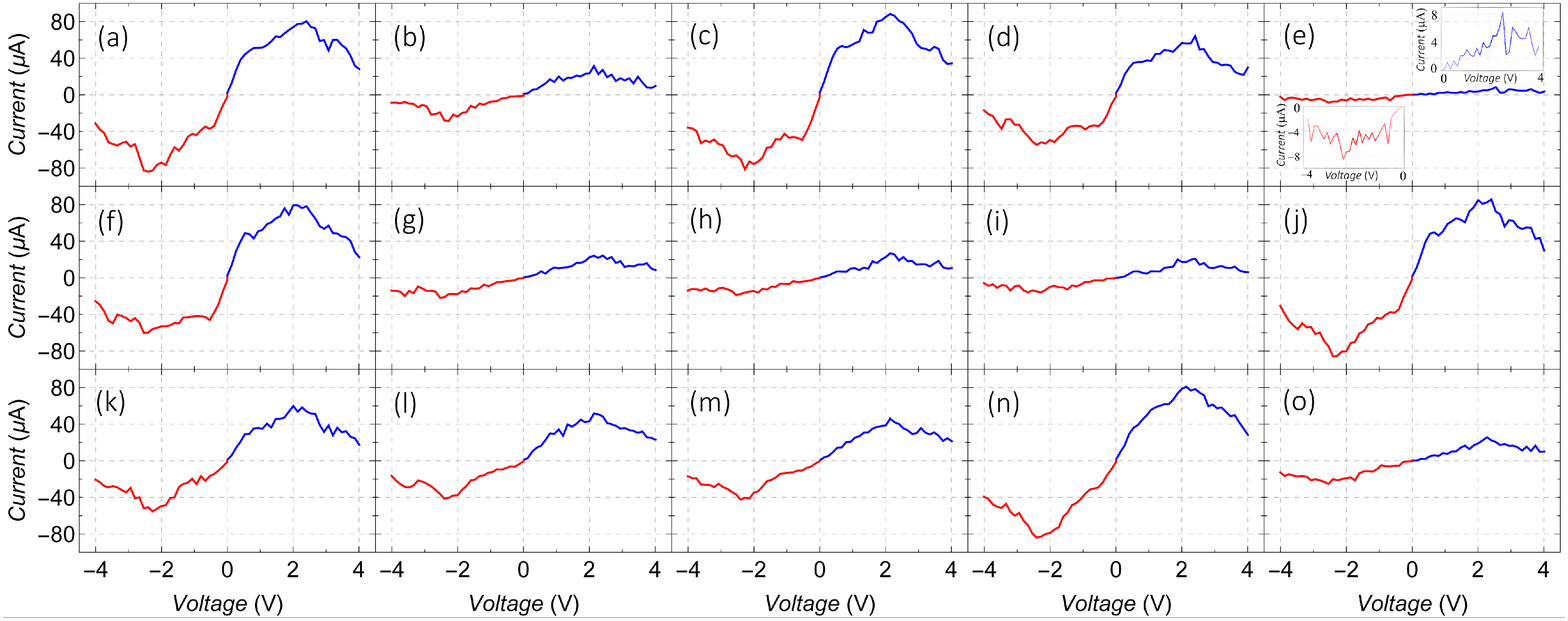}
	\caption{(Color online). Current-voltage relationship corresponding to the nanopores of Fig.~\ref{fig1}. The ordering of the results are similar to that of Fig.~\ref{fig1}. Color conventions are similar to that of Fig.~\ref{fig2}(b). Fermi energy is fixed at $E_F=0.5\,$eV.}
	\label{fig4}
\end{figure*}

\subsection{Pristine ZGNR}

Let us begin with the results for the pristine ZGNR, which is without any kind of nanopore and contains 5382 carbon atoms. The behavior of transmission probability as a function of energy is shown in Fig.~\ref{fig2}(a) for three different voltages, namely, 0.5$\,$V, 1$\,$V, and 1.5$\,$V, and the corresponding results are denoted with violet, orange, and green colors, respectively. For all three voltages, the transmission probabilities are identical to each other for the forward and reverse biased conditions, as the system is symmetric. Moreover, we see that with increasing the bias voltage, the transmission spectrum shows a decreasing nature. This is due to the following reason. In the presence of a finite bias, owing to the developed potential profile across the sample, the site energies become field-dependent. Under such a scenario, the site energies are voltage-dependent and are no longer identical. This is analogous to a conventional disordered lattice, which causes electronic localization. Such kind of localization phenomenon is known as the Wannier-Stark (WS) localization~\cite{loc1,loc2,loc3}.  

To compute the current in the forward bias, the source is attached to the left and the drain at the right sides of the active region, while in the reverse bias case, the source and drain positions are interchanged. We make use of the convention that a current flowing from left (right) to right (left) is a positive (negative) current. As by Eq.~\ref{curr}, the current for a given voltage is computed by the area under the transmission spectrum, the $I$-$V$ characteristics for the two biased conditions are identical to each other as is seen in the inset of Fig.~\ref{fig2}(b). In both the biased conditions, the current tends to drop with voltage around 3$\,$V, indicating a negative differential resistance (NDR) effect as a consequence of the WS localization. Since the magnitudes of the currents at a given voltage are the same in both the reverse and forward biased conditions, the $RR$ becomes zero (as $ \mid I(+V) \mid- \mid I(-V) \mid = 0 $). Thus, even in the presence of a voltage drop across the sample, pristine ZGNR does not exhibit any net current rectification. In order to achieve a finite RR in ZGNR, we must induce asymmetry in the sample by some means, which is accomplished by introducing nanopores in the present work. The corresponding results are discussed as follows.

Now, let us focus on the rectification properties of ZGNRs in the presence of variously structured nanopores, which may provide significant new information and can yield some intriguing findings. As mentioned earlier, a nanopore must be placed asymmetrically relative to the source and drain. If the nanopore is placed at the center of the ZGNR, no current rectification is found, which we have confirmed (not shown here). Thus, the results reported here are pertaining to the situations where the nanopores are created on one side of the ZGNR. To get a comparative analysis, the variously shaped nanopores are created in such a way that the number of removed carbon atoms from the parent lattice (i.e., pristine graphene) is comparable for all the systems as depicted in Fig.~\ref{fig1} and is also listed in Table~\ref{tab1}.

\begin{figure*}[t]
\centering
\includegraphics[width=1\textwidth]{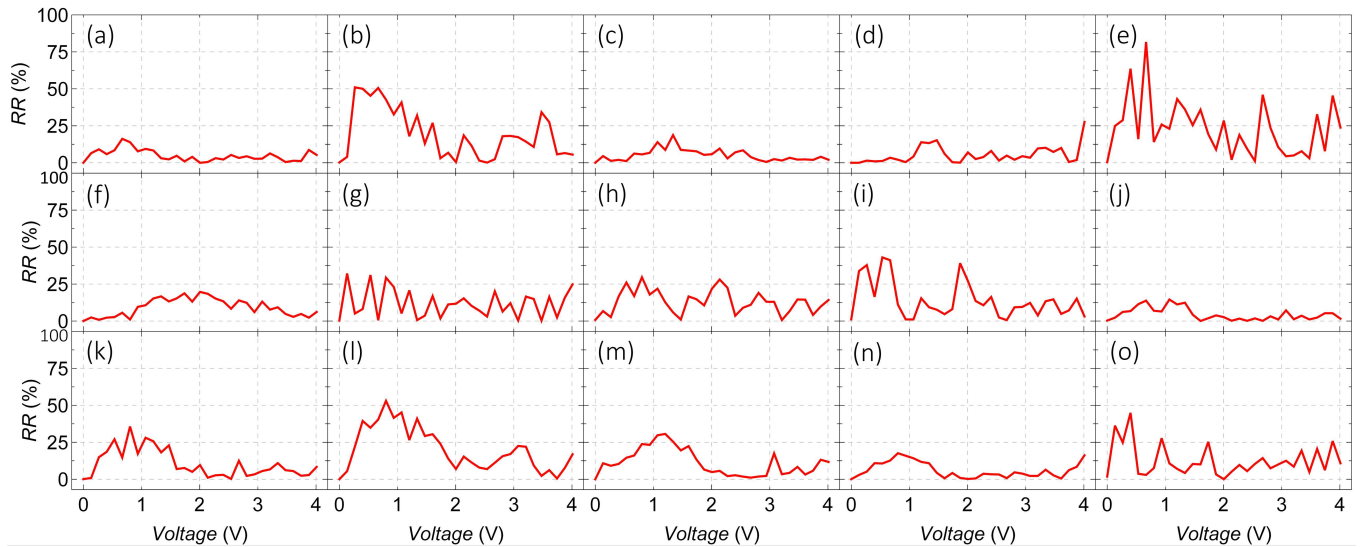}
	\caption{(Color online). Rectification ratio ($RR$) as a function of bias voltage derived from Fig.~\ref{fig4} for all the nanopores. The ordering of the results are similar to that of Fig.~\ref{fig1}.}
	\label{fig5}
\end{figure*}

For all the shapes of the nanopores considered in the present work, the transmission spectrum and the $I$-$V$ plots for two biased conditions are depicted in Figs.~\ref{fig3} and \ref{fig4}, respectively to have a comparative study among all possible shapes of nanopores. As the WS localization is always present for any potential profile, we show the transmission probability as a function of energy at only one particular voltage, namely, at $0.5\,$V, both in the forward and reverse biased conditions. Finally, the degree of rectification $RR$ as a function of voltage is shown in Fig.~\ref{fig5}, which also includes the results of all possible nanopore shapes. We now discuss them one by one.    

\subsection{Circular and elliptical nanopores}
{\it Circular NP}: The circular nanopore with a radius of 5$\,$nm is considered by removing 792 carbon atoms out of 5382 atoms from the pristine ZGNR. Figure~\ref{fig3}(a) depicts the transmission spectrum under two biased conditions in presence of circular NP. The blue line represents the transmission probability for the forward biased condition, whereas the red line represents the transmission probability for the opposite polarity. The transmission probability drops considerably compared to the pristine case due to the lesser accessible conducting pathways for electrons. However, in contrast to the pristine case, the transmission spectra are distinguishable for the two biased conditions, which is completely due to the asymmetry in the conducting channel induced by the circular NP. This ensures a finite current rectification. Careful observation reveals that for a particular Fermi energy, the transmission probability approaches zero for one biased condition, while it is finite for the other biased condition. The currents for the forward (blue line) and reverse biased (red line) conditions in presence of circular NP are represented in Fig.~\ref{fig4}(a). The currents are of the order of $\mu\,$A and different for two biased conditions. It also shows an NDR effect, that is, the value of the current drops significantly beyond the voltages $V=\pm 2\,$V. The corresponding rectification ratio is shown in Fig.~\ref{fig5}(a) as a function of voltage. A finite rectification is observed even at low bias. The highest rectification ratio obtained here is about $16.2\%$ around $1\,$V. However, the rectification ratio in the present case is relatively low.

{\it Elliptical NPs}: For the elliptical NPs, the number of carbon atoms removed from the pristine ZGNR is also very close to that of the circular NP (see Table~\ref{tab1}). Here two elliptical NPs are considered with different orientations. (i) Ellipse-I: the semi-major axis is along the width of ZGNR (Fig.~\ref{fig1}(b)) and (ii) ellipse-II: the semi-major axis is along the length of the ZGNR (Fig.~\ref{fig1}(c)). Ellipse-I NP has a semi-major axis of about $4\,$nm and a semi-minor axis of about $1.5\,$nm, while those are about $3.5\,$nm and $1.7\,$nm for ellipse-II NP. The transmission plots of ellipse-I and II are depicted in Figs.~\ref{fig3}(b) and (c), respectively. In both cases, the characteristic features of the $T$-$E$ spectrum are different for the forward and reverse biased conditions. The transmission probabilities have lower values for ellipse-I than that of ellipse-II, which is due to the lesser available paths for the electrons to flow from the source to the drain. Consequently, the current is also lower for the first kind of elliptical NP (Fig.~\ref{fig4}(b)) than that of the other one (Fig.~\ref{fig4}(c)). Though the current is relatively low for the ellipse-I, the rectification efficiency is quite high (Fig.~\ref{fig5}(b)). The maximum RR is noted about 51$\%$, while it is about only 18.7$\%$ for the second type of elliptical NP (Fig.~\ref{fig5}(c)).

It is now established that asymmetric current transfer for forward and reverse biased conditions can be achieved by drilling NPs asymmetrically with regard to the center of the conducting zone. The significant outcome from the circular and elliptical NPs is that the number of atomic sites removed to produce a nanopore perpendicular to electron transmission is more effective than the number of sites removed along the direction of electron motion to obtain a reasonable RR. The highest RR is obtained for the nanopore in the form of a vertical ellipse, and from this discussion, we can deduce what sort of nanopore should be chosen among similar kinds of geometries and where it should be positioned to obtain the maximum value of current rectification.

\subsection{ Square and rectangular nanopores}
{\it Square NP}: The square-shaped NP is considered here by removing about 14.7$\%$ carbon atoms from the pristine ZGNR and it has a side of $4.5\,$nm. The behavior of the transmission coefficients for the two biased conditions is different from each other as is seen in Fig.~\ref{fig3}(d). The currents for both the biases are of the order of $\mu$A (Fig.~\ref{fig4}(d)). The behavior of current as a function of voltage for the forward bias is different from the reveres bias case. However, the rectification performance is not satisfactory in the given case (Fig.~\ref{fig5}(d)) as the maximum $RR$ is noted about 27.6$\%$ at around $4\,$V.

{\it Rectangular NPs}: Like the case of elliptical NPs, here also we consider two rectangular NPs, having different orientations. The longer side of one rectangular NP is along the width of the ZGNR, which we call rectangle-I (Fig.~\ref{fig1}(e)). The other rectangular NP has the longer side along the length of the ZGNR and is named rectangle-II (Fig.~\ref{fig1}(f)). Rectangle-I has a length about $8.5\,$nm and width about $2.5\,$nm, while those for rectangle-II are about $6.5\,$nm and $3\,$nm, respectively. For both the rectangular NPs, the percentage of removed sites is similar to those previously discussed NPs. The transmission spectrum of rectangle-I shows lower values (Fig.~\ref{fig3}(e)) compared to the rectangle-II case (Fig.~\ref{fig3}(f)) due to a similar reason as discussed earlier for the elliptical NPs. Correspondingly, the current is also quite small for rectangle-I (Fig.~\ref{fig4}(e)) than that for rectangle-II (Fig.~\ref{fig4}(f)). For rectangle-I case, the current is only a few $\mu$A as is seen from the insets of Fig.~\ref{fig4}(e), where the current for positive bias is shown at the top right and that for the negative bias at the bottom left in the same plot. As predicted in the case of elliptical NPs, rectangle-I exhibits a remarkable rectification performance (Fig.~\ref{fig5}(e)), where the maximum $RR$ is observed about 81.7$\%$ at around 0.8$\,$V. On the other hand, for the other rectangular NP, $RR$ is quite low (Fig.~\ref{fig5}(f)), about 19.7$\%$.

\subsection{Triangular nanopores}

A triangular NP can have several possible structures depending upon the arm geometry (zigzag or armchair) and length. Here we consider four different triangular structures: (i) an equilateral triangle with all sides having the armchair shape (AAA: Fig.~\ref{fig1}(g)), (ii) a triangle with two sides having the armchair shape while the other has the zigzag shape (AAZ: Fig.~\ref{fig1}(h)), where all the sides are unequal, (iii) an isosceles triangle with two equal sides having zigzag shapes and the other side having the armchair shape (AZZ: Fig.~\ref{fig1}(i)), and an equilateral triangle with all the sides having the zigzag shape (ZZZ: Fig.~\ref{fig1}(j)). The letter `A' denotes an arm with an armchair shape and `Z' signifies an arm with a zigzag shape. For the triangular NPs with configurations AAA and AAZ, the number of sites removed from the pristine ZGNR is about 14.5$\%$, which is close to the previously discussed NPs, On the other hand, for the other two triangular NPs (AZZ and ZZZ), the deleted number of sites are lower than those for the former two triangular NPs (Table~\ref{tab1}).

Figures~\ref{fig3}(g), (h), (i), and (j) show the behaviors of the transmission probability for the triangular NPs having configurations AAA, AAZ, AZZ, and ZZZ, respectively. The ZZZ triangular NP show larger transmission values compared to the other three NPs. The corresponding current is also higher for the ZZZ NP (Fig.~\ref{fig4}(j)) than other triangular NPs (Figs.~\ref{fig4}(g-i)). In all the $I$-$V$ plots, the NDR effect is present. The rectification ratios are plotted in Figs.~\ref{fig5}(g-j) for the four triangular NPs. We find that $RR$ is lowest for ZZZ triangular nanopore, while the rest of the three NPs exhibit significant values of $RR$. Most importantly, The triangular NP with AZZ configuration has the lowest number of removed sites and yet it displays a large $RR$, about 53$\%$. In this context, it is important to note that, the maximum $RR$ is obtained for the larger sides that are parallel to the width of the ZGNR and the observation is consistent with the previous results for the elliptical and rectangular NPs. The side length parallel to the width of ZGNR for the AAA NP is about 8.5$\,$nm, and that for the AAZ, AZZ, and ZZZ NPs are about 9$\,$nm, 9$\,$nm, and 6.5$\,$nm respectively. 

\begin{center}
\begin{table}[ht]
\caption{The number of carbon atoms ($N_0$) of pristine ZGNR, number of removed sites, percentage of removed sites, and rectification ratio (in $\%$) for all the nanopore geometries considered in the present work.}
	\label{tab1}
	\begin{tabular}{|c|c|c|c|c|} 
		\hline
		\shortstack{Geometry of\\ nanopore} & $N_0$ & \shortstack{No. of sites \\ removed} & \shortstack{$\%$ of sites\\ removed} & RR (in $\%$) \\
		\hline
		None & 5382 & 0 & 0 & 0 \\ \hline
		Circle & 5382 & 792 & 14.7 & 16.2  \\ \hline
		Ellipse-A & 5382 & 790 & 14.7 & {\bf 51.0} \\\hline
		Ellipse-B & 5382 & 794 & 14.7 & 18.7 \\\hline
		Square & 5382 &  790   &  14.7 & 27.6  \\\hline
		Rectangle-I  & 5382 & 798 &   14.8 & \bf {81.7}   \\\hline
		Rectangle-II & 5382 & 793 & 14.8 & 19.7 \\\hline
		Triangle (AAA) & 5382 & 778 & 14.5 & 32.2 \\\hline
		Triangle (AAZ) & 5382 & 781 & 14.5 & 29.6 \\\hline
		Triangle (AZZ) & 5382 & 502 & 9.3 & 43.0 \\\hline
		Triangle (ZZZ) & 5382 & 562 & 10.4 & 14.5 \\\hline
		Arbitrary shape-I & 5382 & 794 & 14.7 & 35.7  \\\hline
		Arbitrary shape-II & 5382 & 792 & 14.7 & {\bf 53.0} \\\hline
		Arbitrary shape-III & 5382 & 797 & 14.8 & 30.7\\\hline
		Arbitrary shape-IV & 5382 & 805 & 14.9 & 17.6  \\\hline
		Random cavity & 5382 & 792 & 14.7 & 44.5 \\
		\hline
	\end{tabular}
\end{table}
\end{center}

\subsection{Arbitrarily shaped nanopores}

So far, we have discussed the rectification properties for different regularly shaped nanopores, out of which the rectangular NP with its longer side along the width of the ZGNR shows the highest $RR$ ($\sim 81.7\%$) and the elliptical NP with its semi-major axis along the width of the ZGNR displays a moderate $RR$ ($\sim 51\%$). Now, it is not always possible to fabricate a perfectly regularly shaped NP experimentally and may have some irregularities in its structure. To include such possibilities, we now consider a few arbitrarily shaped NP structures as shown in the bottom row of Fig.~\ref{fig2}, where the number of removed sites are more or less similar to the previous cases (Table~\ref{tab1}).

The transmission spectra corresponding to the arbitrarily shaped NPs are shown in the bottom row of Fig.~\ref{fig3}, namely is Figs.~\ref{fig3}(k-o). In all the cases, the behavior of the transmission probability as a function of energy for forward bias is different from that in the reverse bias case. In the bottom rows of Figs.~\ref{fig4} and \ref{fig5}, the $I$-$V$ characteristics and rectification ratio for a wide bias window are presented for each case. We notice a maximum rectification ratio of about $53\%$ for the NP with arbitrary shape-II (Fig.~\ref{fig5}(l)). For the random cavity case, we also observe a moderate $RR$ of about $44.5\%$ (Fig.~\ref{fig5}(o)).

\subsection{Effect of nanopore size}

Until now, we have shown a comparative study of several possible NP structures by deleting a similar number of deleted sites as far as the geometry of the NP allows it. 
\begin{figure}[ht]
	{\centering \includegraphics[width=0.4\textwidth]{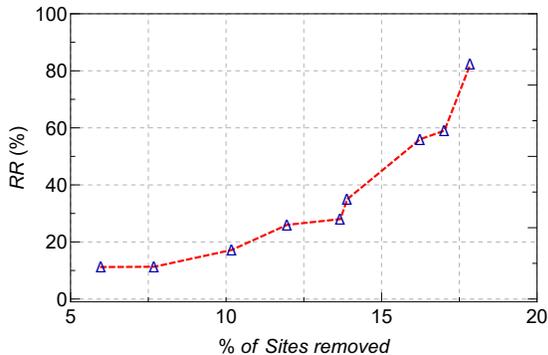}}
	\caption{(Color online). Behavior of $RR$ with the size of the rectangular nanopore (rectangle-I). Fermi energy is fixed at $E_F=0.5\,$eV.}
	\label{fig6}
\end{figure}
It is also important to look at the influence of nanopore size on rectification property. Therefore, to investigate the effect of nanopore size, we consider a rectangular NP (Fig.~\ref{fig1}(e)) for which the maximum $RR$ was obtained. We vary the NP size by removing the atomic sites from the pristine ZGNR (having the identical dimensions as mentioned earlier) keeping the width of the rectangular NP fixed. The percentage of removed sites are varied between $5\%$ to $18\%$. The Fermi energy is fixed at 0.5$\,$eV. The variation of $RR$ with the percentage of sites eliminated is shown in Fig.~\ref{fig6}. We see that the $RR$ increases monotonically with increasing the percentage of removed sites. Thus, we can have different degrees of rectification by suitably adjusting the size of the NP.

\subsection{Choice of Fermi energy}

\begin{figure}[ht]
	{\centering \includegraphics[width=0.4\textwidth]{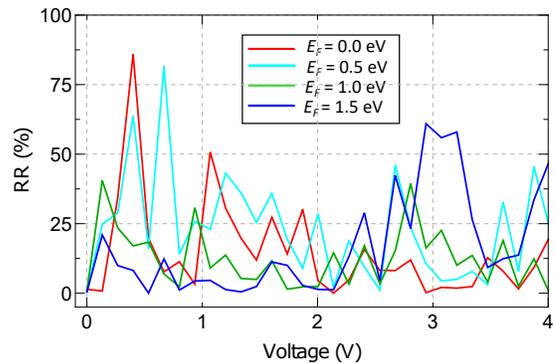}}
	\caption{(Color online). Variation of rectification ratio ($RR$) with voltage for four different Fermi energies, namely $E_F=0\,$eV (red), 0.5$\,$eV (cyan), 1$\,$eV (green), and 1.5$\,$eV (blue). The nanopore shape is rectangular-I.}
	\label{fig7}
\end{figure}

In all the $I$-$V$ characteristics discussed so far, the Fermi energy was fixed at 0.5$\,$eV. To see the precise role of Fermi energy on the rectification property, we study the behavior of $RR$ as a function of bias voltage for four different Fermi energies as shown in Fig.~\ref{fig7}. The chosen four Fermi energy values are $E_F = 0$, 0.5, 1, and 1.5, and the corresponding results are denoted with red, cyan, green, and blue colors, respectively. The NP structure considered here is the rectangular NP, namely the shape rectangle-I (Fig.~\ref{fig1}(e)). 
For $E_F=0$ and 0.5, more than 80$\%$ $RR$ is obtained at low bias voltages. We also observe more than $60\%$ $RR$ for $E_F=1.5$ at around 3$\,$V. Thus it is apparent that the choice of Fermi energy plays a crucial role to achieve a significant current rectification at the desired voltage. This is because the junction current considerably gets modified with the change of Fermi energy depending on the available transmission peaks across it.

\subsection{Role of potential profile}

Under the application of a non-zero bias between the source and drain, 
\begin{figure}[ht]
	{\centering \includegraphics[width=0.4\textwidth]{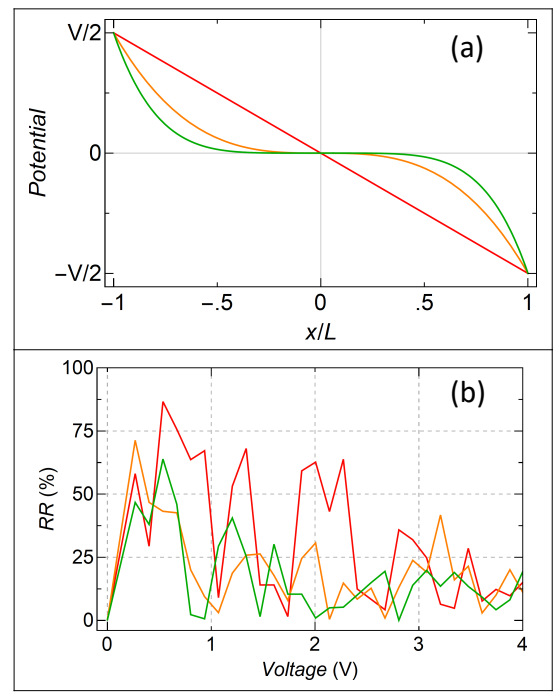}}
	\caption{(Color online). (a) The spatial distribution of field-dependent site energies on a pristine ZGNR for three different electrostatic potential profiles. The linear profile is denoted with red color and two non-linear potential profiles are denoted with orange and green colors (less steep). (b) $RR$ (in $\%$) as a function of bias voltage $V$ for the three electrostatic potential profiles using a rectangular nanopore (rectangle-I). Fermi energy is fixed at $E_F=0.5\,$eV. }
	\label{fig8}
\end{figure}
an electrostatic potential profile is developed across the system. Consequently, the site energies become voltage-dependent, as already mentioned earlier. The real variant of this potential profile is extremely difficult to implement since it entails a complete many-body issue. The most prevalent form is linear dependency, which we have addressed in our work so far. 
However additional forms related to electron screening in other materials should also be considered. Therefore, to get a comprehensive understanding of the bare and screened electric field profiles, we take two different non-linear potential profiles~\cite{lyuan} including the linear case and the corresponding spatial distribution of on-site potentials are shown in Fig.~\ref{fig8}(a). The analytical form for the linear case is $\epsilon_i(V) = V\left(x/L\right)$ (shown by red curve), where $2L$ is the length of the ZGNR. For one of the non-linear cases, it is $\epsilon_i(V) = V\left(x^3/L\right)$ (shown by orange curve) and for the other case, the potential profile has the form $\epsilon_i(V) = V\left(x^5/L\right)$ (shown by green curve).

Here also we consider the same rectangular nanopore (rectangle-I), keeping all the previously specified parameters unchanged. The result for linear potential profile is already studied 
earlier and it is included for the sake of comparison. One of the non-linear profiles exhibits a good rectification behavior, as denoted with orange color, where $RR$ is more than $70\%$ at very low voltage. The maximum $RR$ for the other potential profile also shows a moderate value, about $62\%$ denoted with the green color. Overall, the envelope of the $RR$-$V$ curves for the three different potential profiles reveal that more steepness in the potential profile leads to greater $RR$.

\section{conclusion}

To summarise, in the present work, we have given a new scheme to achieve rectification in ZGNR with various regularly and irregularly shaped nanopores, such as circles, ellipses, rectangles, triangles, and a few arbitrarily formed geometries, by placing them in such a way that the system becomes physically asymmetric. The key findings of our analysis are as follows : 

$\bullet$ All the NPs considered in the present work exhibit rectification behavior.

$\bullet$ Negative differential resistance is observed in all the cases as a consequence of Wannier-Stark localization.

$\bullet$ The maximum rectification ratio is obtained for a specific type of rectangular nanopore, which is more than $80\%$ at a low bias. 

$\bullet$ The rectification ratio can be controlled by varying the height of the rectangular nanopore and that applies to the other shapes also.

$\bullet$ The choice of Fermi energy plays a crucial role in achieving a considerable rectification of the current.

$\bullet$ The steepness of the potential profile plays important role in getting a higher rectification ratio. The higher the steepness, the larger the efficiency.

We thus conclude this discussion by stating that ZGNR, when equipped with an asymmetrically positioned nanopore, can function as an efficient rectifier with a high rectification ratio. A similar kind of outcome is also very likely to be expected for the AGNR case. Owing to the simplicity of the proposal, the possibility of drilling nanopores with different shapes and sizes experimentally, the given prescription can also be explored in other graphene derivatives, which may find efficient rectification signatures.  

\section*{acknowledgement}

JM is thankful for the financial support provided by CSIR, India (09/093(0185)/2019-EMR-I) for conducting his research fellowship. SKM thankfully acknowledges the financial support of the SERB, DST, Government of India (Project File Number: EMR/2017/000504).


\begin{thebibliography}{99}
	
	\bibitem{konenkamp85} R. Könenkamp, R. C. Word, and C. Schlegel, App. Phys. Lett. \textbf{85}, 6004 (2004).
	
	\bibitem{aviram29} A. Aviram and M. A. Ratner, Chem. Phys. Lett. \textbf{29}, 277 (1974).
	
	\bibitem{dhirani106} A. Dhirani, P. -H. Lin, P. Guyot-Sionnest, R. W. Zehner, and L. R. Sita, J. Chem. Phys. \textbf{106}, 5249 (1997).
		
	\bibitem{zhou71} C. Zhou, M. R. Deshpande, M. A. Reed, L. Jones II, and J. M. Tour, Appl. Phys. Lett. \textbf{71}, 611 (1997).
		
	\bibitem{frantti59} J. Frantti, V. Lantto, S. Nishio, and M. Kakihana, Phys. Rev. B \textbf{59}, 12 (1999).
		
	\bibitem{metzger32} R. M. Metzger, Acc. Chem. Res. \textbf{32}, 950 (1999).
		
	\bibitem{taylor89} J. Taylor, M. Brandbyge, and K. Stokbro, Phys. Rev. Lett. \textbf{89}, 138301 (2002).
		
	\bibitem{ashwell126} G. J. Ashwell, W. D. Tyrrell, and A. J. Whittam, J. Am. Chem. Soc. \textbf{126}, 7102 (2004).
		
	\bibitem{perez1} I. D\'{i}ez-P\'{e}rez, J. Hihath, Y. Lee, L. Yu, L. Adamska, M. A. Kozhushner, I. I. Oleynik, and N. Tao, Nat. Chem. \textbf{1}, 635 (2009).
		
	\bibitem{nijhuis132} C. A. Nijhuis, W. F. Reus, and G. M. Whitesides, J. Am. Chem. Soc. \textbf{132}, 18386 (2010).
		
	\bibitem{yee5} S. K. Yee et al., ACS Nano \textbf{5}, 9256 (2011).
		
	\bibitem{batra13} A. Batra et al., Nano Lett. \textbf{13}, 6233 (2013).
		
	\bibitem{yoon136} H. J. Yoon et al., J. Am. Chem. Soc. \textbf{136}, 17155 (2014).
		
	\bibitem{wang141} K. Wang, J. Zhou, J. M. Hamill, and B. Xu, J. Chem. Phys. \textbf{141}, 054712 (2014)
		
	\bibitem{mujica104} V. Mujica, M. Kemp, A. Roitberg, and M. Ratner, J. Chem. Phys. \textbf{104}, 7296 (1996).
		
	\bibitem{mujica281} V. Mujica, M. A. Ratner, and A. Nitzan, Chem. Phys. \textbf{281}, 147 (2002).
		
	\bibitem{krzeminski64} C. Krzeminski, C. Delerue, G. Allan, D. Vuillaume, and R. M. Metzger, Phys. Rev. B \textbf{64}, 085405 (2001).
		
	\bibitem{kornilovitch66} P. E. Kornilovitch, A. M. Bratkovsky, and R. S. Williams, Phys. Rev. B \textbf{66}, 165436 (2002).

		
	\bibitem{zahid70} F. Zahid, A. W. Ghosh, M. Paulsson, E. Polizzi, and S. Datta, Phys. Rev. B \textbf{70}, 245317 (2004).
		
	\bibitem{metzger212} R. M. Metzger, Macromol. Symp. \textbf{212}, 63 (2004).
		
	\bibitem{dalgleish73} H. Dalgleish and G. Kirczenow, Phys. Rev. B \textbf{73}, 245431 (2006).

	\bibitem{saha93} M. Saha and S. K. Maiti, Physica E \textbf{93}, 275 (2017).
		
	\bibitem{patra484} M. Patra and S. K. Maiti, Jour. Magnetism and Magnetic Material \textbf{484}, 408 (2019).

	\bibitem{saha52} M. Saha, S. K. Maiti, J. Phys. D: Appl. Phys. {\bf 52}, 465304 (2019).

	\bibitem{melbing} M. Elbing, R. Ochs, M. Koentopp, M. Fischer, C. v. H\"{a}nisch, F. Weigend, F. Evers, H. B. Weber, and M. Mayor, Natl. Acad. Sci. USA {\bf 102}, 8815 (2005).

	\bibitem{span} S. Pan, Q. Fu, T. Huang, A. Zhao, B. Wang, Y. Luo, J. Yang, and J. Hou, Proc. Natl. Acad. Sci. USA {\bf 36}, 15259 (2009).
			
	\bibitem{guo86} A. Guo and Q.-F. Sun, Phys. Rev. B \textbf{86}, 115441 (2012).
		
	\bibitem{guo8} C. Guo \textit{et al.}, Nature Chem. \textbf{8}, 484 (2016).

	\bibitem{rev-mol} D. Xiang, X. Wang, C. Jia, T. Lee, and X. Guo, Chem. Rev. {\bf 116}, 4318 (2016).
		
	\bibitem{neto81} A. H. C. Neto \textit{et al.}, Rev. Mod. Phys. \textbf{81}, 109 (2009). 
		
	\bibitem{liao15} L. Liao and X. Duan, Mater Today \textbf{15}, 328 (2012). 
		
	\bibitem{schwierz101} F. Schwierz, Proc. IEEE \textbf{101}, 1567 (2013).

	\bibitem{han-prl} M. Y. Han, B. \"{O}zyilmaz, Y. Zhang, and 
P. Kim, Phys. Rev. Lett. {\bf 98}, 206805 (2007).

	\bibitem{dwei} D. Wei, L. Xie, K. K. Lee, Z. Hu, S. Tan, W. Chen, C. H. Sow, K. Chen, Y. Liu, and A. T. S. Wee, Nat. Commun. {\bf 4}, 1374 (2013).

	\bibitem{xhzang} X. -H. Zhang, X.  -F. Li, L. -L. Wang, L. Xu, K. -W. Luo, Appl. Phys. Lett. {\bf 104}, 103107 (2014).

	\bibitem{fwang} Z. F. Wang, Q. Li, Q. W. Shi, X. Wang, J. G. Hou, H. Zheng, and J. Chen, Appl. Phys. Lett. {\bf 92}, 133119 (2008).

	\bibitem{xfli} X. -F. Li, L. -L. Wang, K. -Q. Chen, and Y. Luo, J. Phys. Chem. C {\bf 115}, 12616 (2011).

	\bibitem{jzeng} J. Zeng, K.-Q. Chen, J. He, X.-J. Zhang, C.Q. Sun, J. Phys. Chem. C {\bf 115}, 25072 (2011).

	\bibitem{xlji} X. -L. Ji, Z. Xie, X. Zuo, G. -P. Zhang, Z. -L. Li, C. -K. Wang, Phys. Lett. A {\bf 380}, 3198 (2016).

	\bibitem{tchen} T. Chen, X. -F. Li, L. -L. Wang, K. -W. Luo, and L. Xu, J. Appl. Phys. {\bf 116}, 013702 (2014).

	\bibitem{lxie} L. Xie, S.  -Z. Chen, W. -X. Zhou, K. -Q. Chen, Org. Electron. {\bf 46}, 150 (2017).

	\bibitem{xh-chem} X. -H. Zhang, S. -J. Liu, L. Tian, Q. Wan, A. -M. Hu, X. -F. Li, Chem. Phys. {\bf 546}, 111140 (2021).

	\bibitem{eszam} E. Zaminpayma, P. Nayebi, and M. Emami-Razavi, Nanotechnol. {\bf 32} 205204 (2021).


		
	
	\bibitem{bai5} J. W. Bai et al., Nat. Nanotechnol. \textbf{5}, 190 (2010).
	
	\bibitem{tada107} K. Tada et al., Phys. Rev. Lett. \textbf{107}, 217203 (2011).
	
	\bibitem{Fischbein93} M. D. Fischbein and M. Drndi´c, Appl. Phys. Lett. \textbf{93}, 11 (2008).
	
	\bibitem{zhang6} Q. Zhang et al., Nano Res. \textbf{6}, 478 (2013).
	
	\bibitem{safron8} N. S. Safron, M. Kim, P. Gopalan, and M. S. Arnold, Advanced Materials 24, \textbf{8} (2012).
	
	\bibitem{kalhor114} N. Kalhor, S. A. Boden, and H. Mizuta, Microelectron. Eng. \textbf{114}, 70 (2014).
	
	\bibitem{abbas8} A. N. Abbas et al., Acs Nano \textbf{8}, 1538 (2014)
		
	\bibitem{pleutin118} S. Pleutin, H. Grabert, G. L. Ingold, and A. Nitzan, J. Chem. Phys. \textbf{118}, 3756 (2003).
	
	\bibitem{maiti377} S. K. Maiti and A. Nitzan, Phys. Lett. A \textbf{377}, 1205 (2013).

	\bibitem{cguo} C. Guo, K. Wang, E. Zerah-Harush, J. Hamill, B. Wang, Y. Dubi, and B. Xu, Nat. Chem. {\bf 8}, 484 (2016). 

	\bibitem{msaha} M. Saha and S. K. Maiti, J. Phys. D: Appl. Phys. {\bf 52}, 465304 (2019).

	\bibitem{kwant} C. W. Groth, M. Wimmer, A. R. Akhmerov, and
  X. Waintal, New J. Phys. {\bf 16}, 063065 (2014).

\bibitem{etms} S. Datta, {\it Electronic Transport in Mesoscopic Systems},
Cambridge University Press, Cambridge, 1995.

\bibitem{qtat} S. Datta, {\it Quantum Transport: Atom to Transistor},
Cambridge University Press, Cambridge, 2005.

\bibitem{Fisher-Lee} D. S. Fisher and P. A. Lee, Phys. Rev. B
  {\bf{23}}, 6851 (1981).

	\bibitem{loc1} G. H. Wannier, Phys. Rev. {\bf 117}, 432 (1960).
	 
	\bibitem{loc2} J. R. Borysowicz, Phys. Lett. A {\bf 231}, 240 (1997).

	\bibitem{loc3} N. Zekri, M. Schreiber, R. Ouasti, R. Bouamrane, and A. Brezini, Z. Phys. B {\bf 99}, 381 (1996).

   \bibitem{lyuan} L. Yuan, N. Nerngchamnong, L. Cao, H. Hamoudi, E. del Barco, M. Roemer, R. K. Sriramula, D. Thompson, and C. A. Nijhuis, Nat. Commun. {\bf 6}, 6324 (2015).
	
	


	
\end{thebibliography}
\end{document}